\documentclass[pdflatex,sn-mathphys-num,twocolumn]{sn-jnl}

\usepackage{graphicx}%
\usepackage{multirow}%
\usepackage{amsmath,amssymb,amsfonts}%
\usepackage{amsthm}%
\usepackage{mathrsfs}%
\usepackage[title]{appendix}%
\usepackage{xcolor}%
\usepackage{textcomp}%
\usepackage{manyfoot}%
\usepackage{booktabs}%
\usepackage{algorithm}%
\usepackage{algorithmicx}%
\usepackage{algpseudocode}%
\usepackage{listings}%

\theoremstyle{thmstyleone}%
\theoremstyle{thmstyletwo}%

\theoremstyle{thmstylethree}%

\begin{document}

\title[Strongly resonant polarization dynamics of pulse trains]{Strongly resonant polarization dynamics of pulse trains in a spin-flip model for an excitable microlaser with delayed self-feedback}


\author*[1]{\fnm{Stefan} \sur{Ruschel}}\email{s.ruschel@leeds.ac.uk}

\author[2]{\fnm{Sylvain} \sur{Barbay}}\email{sylvain.barbay@universite-paris-saclay.fr}
\equalcont{These authors contributed equally to this work.}

\author[3]{\fnm{Neil G. R.} \sur{Broderick}}\email{n.broderick@auckland.ac.nz}
\equalcont{These authors contributed equally to this work.}

\author[4]{\fnm{Bernd} \sur{Krauskopf}}\email{b.krauskopf@auckland.ac.nz}
\equalcont{These authors contributed equally to this work.}

\affil*[1]{\orgdiv{School of Mathematics}, \orgname{University of Leeds}, \orgaddress{\street{Woodhouse Lane}, \city{Leeds}, \postcode{LS2 9JT}, \state{State}, \country{United Kingdom}}}

\affil[2]{\orgdiv{Centre de Nanosciences et de Nanotechnologies}, \orgname{Universit{\'e} Paris-Saclay}, \orgaddress{\street{10 Bd Thomas Gobert}, \city{Palaiseau}, \postcode{91120}, \state{State}, \country{France}}}

\affil[3]{\orgdiv{Department of Physics and Dodd-Walls  Centre for Photonic and Quantum Technologies}, \orgname{The University of Auckland}, \orgaddress{\street{Private Bag 92019}, \city{Auckland}, \postcode{1142}, \state{}, \country{New Zealand}}}

\affil[4]{\orgdiv{Department of Mathematics and Dodd-Walls  Centre for Photonic and Quantum Technologies}, \orgname{The University of Auckland}, \orgaddress{\street{Private Bag 92019}, \city{Auckland}, \postcode{1142}, \state{}, \country{New Zealand}}}


\abstract{Motivated by recent experimental observations concerning the polarization dynamics in an excitable microlaser with saturable absorber coupled to an external feedback mirror reported in [Ruschel et al. (2025) Opt. Lett. \textbf{50}(8) 2618], we propose here an in-depth theoretical investigation of the locked dynamics of regenerative vectorial pulse trains that this system produces. We perform a numerical bifurcation analysis of self-sustained pulse trains of a corresponding spin-flip model with delayed feedback. Its focus is on strongly resonant regimes, where the modulation of the peak intensities of polarized regenerative pulse trains locks to 2, 3 and 4 times the pulse regeneration time. Specifically, we identify points of strongly resonant rotation numbers on curves of torus bifurcations in the parameter plane of the amplitude and phase anisotropy parameters, and continue emerging curves of fold and period-doubling bifurcations to identify locking regions. In this way, we clarify where pulse trains show strongly resonant polarization dynamics and highlight that even weak polarization coupling in delay-coupled microlasers creates considerable dynamical richness. These results may have impact on the design of future  coupled microlasers systems for neuroinspired on-chip computing, where the polarization of an excitable pulse can be used to encode or process information.}




\maketitle



\section{Introduction}
\label{sec:intro}

The capacity of optical devices to generate and regenerate coherent pulses of light is key to a wide range of applications, from optical communication to neuromorphic information processing \cite{McMahonNRP23, Shastri2021, brunner2025roadmap}. Indeed, modern photonic systems are increasingly being designed to mimic the spike-based signaling between biological neurons, with applications ranging from ultrafast signal processing to neuroinspired computing. Among the most promising candidates for such `optical neurons' are microlasers with integrated saturable absorber sections that facilitate excitability and the ability to communicate via optical pulses when coupled \cite{Pammi2019}. To further close the performance gap between neuroinspired photonics and the brain, it has been suggested that polarized pulses (oscillating in a specific direction) could be used to realize not only excitatory (spike reverberation) but also inhibibtory coupling (spike cancellation). 

Polarization describes the orientation of the electric field vector in the plane perpendicular to the direction of propagation of the light wave, and it constitutes an important degree of freedom in pulsing optical systems \cite{Zhao22}. For linear polarization, this vector oscillates along a fixed direction but, in general, it can trace out circular or elliptical paths over time; polarization is typically expressed as a sum of the linear component vectors spanning the plane, or in terms of a circular basis. Cross-coupling between the polarization components introduces rich nonlinear dynamics, including vectorial chaos in VCSELs \cite{ScirePRL03}, vectorial self-organization in mode-locked fiber lasers \cite{KrupaO17}, and symmetry-breaking or breathing vectorial solitons in Kerr and fiber laser resonators \cite{XuNC21, XuOL22, HillCP24, LuoOE20, HuangAPN23}. Orthogonally polarized frequency combs have been predicted numerically in mode-locked VCSELs with saturable absorbers (SAs) \cite{VladimirovOL19}, and vectorial dissipative solitons have been observed experimentally in VCSELs with polarization-selective feedback and cross-polarized reinjection \cite{MarconiNP15}, including coexistence of antiphase bright and dark solitons in orthogonal components.

Here, we focus on regenerative self-pulsation in semiconductor micropillar lasers (MLs) with integrated SAs and delayed optical feedback \cite{GarbinNC15}. The high-reflectivity vertical microvavity of these devices allows laser emission perpendicular to the surface. Polarization selection in MLs possessing circularly symmetric transverse boundaries is mainly influenced by material anisotropies --- such as crystallographic axes orientation --- leading to birefringence and dichroism that couple the orthogonal linear polarization components.  Experimentally, MLs have been demonstrated as building blocks for brain-inspired photonic systems \cite{Pammi2019}, supporting inhibitory neural dynamics \cite{ZhangOL19} and XOR logic operations \cite{XiangOL20} via polarization control. A solitary ML with SA section can emit sustained intensity and polarization pulsations when operated above laser threshold \cite{ScireOL02}; on the other hand, just below the laser threshold this optical system is in the excitable regime and emits calibrated light pulses in response to sufficiently large (above threshold) perturbations \cite{Barbay2011, Selmi2014}. When coupled via delayed optical feedback, an excited pulse can regenerate after each feedback roundtrip  \cite{Terrien2017,Terrien2017a,Terrien2018,Terrien2019,TerrienChaos23,TerrienPRE21, Ruschel2020limits}, forming a train of pulses that are often referred to as temporal dissipative solitons \cite{Yanchuk2019,giraldo2023pulse}. This regime enables optical pulse manipulation --- writing, erasing, and tweezing --- on short timescales for use as optical buffer memories or arbitrary pulse generators \cite{Terrien2018a, PammiThese21}; however, it was shown that on longer timescales, any initial temporal pattern converges to one of several possible stable periodic regimes \cite{Terrien2019,TerrienChaos23, TerrienPRE21}. 

Vectorial regenerative breathers in an excitable ML with delayed self-feedback have been found numerically and experimentally \cite{ruschel2025regenerative}, and they are characterized by an oscillation of the linear polarization angle of successive pulses while the total peak intensity remains nearly constant. This polarization mode competition has intriguing potential applications in photonic neuromorphic circuits. In particular, polarization enables dual-channel encoding of optical spikes and the implementation of excitatory or inhibitory interactions in coupled lasers.   It has been shown numerically  that polarization-mode competition in a VCSEL with a saturable absorber can serve as an all-optical inhibitory mechanism for spiking pulses \cite{ZhangOL19}.  Similarly, by treating the two orthogonal polarizations as separate channels, a single laser can encode and process two streams of spikes simultaneously.  Thus, polarization-resolved dynamics provide a way to realize more complex neural-like behavior (including XOR logic and other computations) in a single device.  These ideas motivate our interest in controlled polarization dynamics: locked, repeatable polarization patterns could be used to encode information in future photonic neurons, or to devise controllable sources of well-defined polarized optical pulse trains. Such locking behavior can also be observed in other setups, for example, in mode-locking ring lasers \cite{wu2025unveiling}.

In this work, we study in detail the polarization locking dynamics of regenerative pulse trains in a corresponding spin-flip model for a ML with a saturable absorber and delayed optical feedback. Specifically, we identify regions in parameter space with locked periodic orbits with periods of approximately 2, 3 and 4 times the roundtrip time. We map out how these locking regimes depend on two important parameters describing the linear polarization anisotropies (dichroism and birefringence) of the cavity.  In particular, we locate codimension-two strong resonance points on torus bifurcation curves. Via the numerical continuation of periodic orbits and their fold and period-doubling bifurcations, we then find the boundaries in the anisotropy parameter plane of the regions where the different locking ratios can be observed. These results considerably extend the limited numerical bifurcation analysis in \cite{ruschel2025regenerative}, which considered only torus bifurcations that mark the onset of peak pulse modulation. 

The paper is organized as follows. In Sec.~\ref{sec:model} we introduce the model equations and discuss its relevant parameter ranges and symmetries. Section~\ref{sec:2ParBifAna} contains the results of our numerical bifurcation analysis of regenerative pulse trains; here, we present the bifurcation diagram in the anisotropy parameter plane of linear dichroism $\varepsilon_a$ and birefringence $\varepsilon_p$, and then discuss the cases of $1\!:\!2$, $1\!:\!3$, and $1\!:\!4$ resonance one-by-one. In Sec.~\ref{sec:concl} we draw conclusion and point out some directions for future research.

\section{Mathematical model, its symmetries and pulse trains}
\label{sec:model}

We consider an extension of the Spin-Flip Model (SFM) for a VCSEL with SA \cite{ScireOL02} that includes optical delayed feedback.  The SFM describes the time-evolution of light-matter interactions between two circular polarization components $F_{\pm}(t)\in\mathbb{C}$ of the complex electric field, with real-valued gain and saturable absorber variables describing the corresponding charge carrier reservoirs. The resulting delay differential equation \cite{Hale1993} model reads
\begin{align}
2 F^\prime_{\pm} =& \left[(1+i\alpha)G_{\pm}
-(1+i\beta)Q_{\pm}-1\right]F_{\pm} \nonumber\\
&
- (\varepsilon_a + i\varepsilon_p) F_{\mp} 
+\kappa e^{i\psi}F_{\pm}(\cdot -\tau),\label{eq:F-def}\\
G^\prime_{\pm} =& \gamma_{G}(A-G_{\pm}(t)(1+|F_{\pm}|^{2})) -\delta_G(G_{\pm}-G_{\mp}),\label{eq:G-def}\\
Q^\prime_{\pm} =& \gamma_{Q}(B-Q_{\pm}(1+a|F_{\pm}|^{2}))
-\delta_Q(Q_{\pm}-Q_{\mp}),
\label{eq:Q-def}
\end{align}
where we use the convention $F_\pm(\cdot - \tau)$ to denote the time shift by time $\tau$, that is, we define $F_\pm(\cdot - \tau)(t)=F_\pm(t - \tau)$ for all $t$. Moreover, $G_\pm$ and $Q_\pm$ represent the population inversion of the gain and loss reservoirs of carriers coupling to the left and right circularly polarized components of the electric field, and time $t$ is rescaled to the photon lifetime in the laser cavity, which is of the order of $1.5$ ps. The parameters $\alpha$ and $\beta$ are the linewidth enhancement factors in the gain and SA regions, and $\gamma_{G,Q}$ are the corresponding relaxation rates, which are small parameters giving rise to slow-fast type dynamics. The parameter $A$ is the pump parameter of the gain region, $B$ is the linear losses parameter of the SA region, $a$ is the saturation parameter, and $\delta_{G,Q}$ are the reduced spin-flip rates in the gain and absorber regions, such that $\delta_{G,Q}=1/2(\gamma_{s_{G,Q}}-\gamma_{G,Q}$) with the spin flip rates $\gamma_{s_{G,Q}}$. The parameters of the optical feedback are the roundtrip time delay $\tau$,  the feedback strength $\kappa$, and the feedback phase difference $\psi$ between the time-delayed and the instantaneous fields.

Polarization anisotropy is represented in Eqs.~\eqref{eq:F-def}--\eqref{eq:Q-def} by the linear dichroism $\varepsilon_a$ and birefringence $\varepsilon_p$, which depend not only on the materials used to fabricate the laser but may also vary as a result of crystal mechanical stress; therefore, their values may vary even for nearby devices on a wafer, due to fabrication-induced anisotropies. It is known, however, that in VCSELs usually the linear dichroism is small whereas birefringence is substantially larger, and this explains the mainly linear polarization emission of these devices \cite{MartinRegaladoIEEEJQE97,ExterPRL98,VanDerSande2006,Panayotov2013}. Note that system \eqref{eq:F-def}--\eqref{eq:Q-def} is invariant under a phase shift over $\pi$ (multiplication by $-1$) applied to only one of the two fields $F_{\pm}$ and changing simulaneously the signs of both $\varepsilon_a$ and $\varepsilon_p$; hence, we may restrict our exploration in the $(\varepsilon_a,\varepsilon_p)$-plane and model analysis to positive values of $\varepsilon_p$.

System \eqref{eq:F-def}--\eqref{eq:Q-def} possesses other symmetries that guide our analysis and aid in the interpretation of its solutions. First, the system is $\mathbb{S}^{1}$-equivariant: the transformation $F_{\pm} \mapsto e^{i\phi}F_{\pm}$, for any fixed phase shift $\phi \in \mathbb{R}$, leaves the equations unchanged. This reflects a common co-rotating frame, where the overall optical phase can be shifted without affecting the system's dynamics. This symmetry allows us to consider intensity and relative phase dynamics without loss of generality.

What is more, system \eqref{eq:F-def}--\eqref{eq:Q-def} is $\mathbb{Z}_{2}$-equivariant, namely swapping the components under the involution $\eta$ with  $\eta(F_{\pm}, G_{\pm}, Q_{\pm}) = (F_{\mp}, G_{\mp}, Q_{\mp})$ does not change \eqref{eq:F-def}--\eqref{eq:Q-def}. In addition, the system supports periodic solutions of temporal period $T$ with spatio-temporal symmetry $(F_{\pm}(t), G_{\pm}(t), Q_{\pm}(t)) = (F_{\mp}(t-T/2), G_{\mp}(t-T/2), Q_{\mp}(t-T/2))$, given by applying the transformation $\eta$ and a time-shift over $T/2$ \cite{golubitsky2003symmetry}. To study this in more detail, it is convenient to introduce the (scaled) longitudinal and transverse coordinates 
\begin{align*}
F_x(t) &= (F_+(t) + F_-(t))/\sqrt{2}, \\
D_1(t) &= (G_+(t)+G_-(t))/2, \\
D_2(t) &= (Q_+(t) + Q_-(t))/2, \ \text{and} \\[1mm]
F_y(t) &= i(F_+(t) - F_-(t))/\sqrt{2}, \\
d_1(t) &= (G_+(t) - G_-(t))/2, \\  
d_2(t) &= (Q_+(t) + Q_-(t))/2,
\end{align*}
defined with respect to the fixed point subspace of the symmetry $\eta$, given by $(F_{\pm}, G_{\pm}, Q_{\pm}) = (F_{\mp}, G_{\mp}, Q_{\mp})$. These variables satisfy the set of delay differential equations
\begin{strip}
\begin{align}
2 F_x^\prime & =\left((1+i\alpha)D_{1}-(1+i\beta)D_{2}-1\right)F_{x}+i\left((1+i\alpha)d_{1}-(1+i\beta)d_{2}\right)F_{y}-\left(\varepsilon_{a}+i\varepsilon_{p}\right)F_{x}+\kappa e^{i\psi}F_{x}(\cdot-\tau)\label{eq:Fx}\\
2 F_y^\prime & =\left((1+i\alpha)D_{1}-(1+i\beta)D_{2}-1\right)F_{y}-i\left((1+i\alpha)d_{1}-(1+i\beta)d_{2}\right)F_{x}+\left(\varepsilon_{a}+i\varepsilon_{p}\right)F_{y}+\kappa e^{i\psi}F_{y}(\cdot-\tau)\\
D_{1,2}^\prime & =\gamma_{1,2}\left[\mu_{1,2}-D_{1,2}\left(1+\frac{1}{2}a_{1,2}\left(\left|F_{x}-iF_{y}\right|^{2}+\left|F_{x}+iF_{y}\right|^{2}\right)\right)-\frac{1}{2}a_{1,2} d_{1,2}\left(\left|F_{x}-iF_{y}\right|^{2}-\left|F_{x}+iF_{y}\right|^{2}\right)\right]\\
d_{1,2}^\prime & =-\gamma_{s1,2}d_{1,2}-\gamma_{1,2}\left[\frac{1}{2}a_{1,2}D_{1,2}\left(\left|F_{x}-iF_{y}\right|^{2}-\left|F_{x}+iF_{y}\right|^{2}\right)+\frac{1}{2}a_{1,2}d_{1,2}\left(\left|F_{x}-iF_{y}\right|^{2}+\left|F_{x}+iF_{y}\right|^{2}\right)\right]\label{eq:d12},
\end{align}
\end{strip}
where $\mu_1 = A, \mu_2 = B$, $a_1=1$, and $a_2=a$. Note that the variables carry physical meaning: $F_{x,y}$ correspond to the linear components of the electric field vector in the $(x,y)$-plane perpendicular to the direction of propagation, $D_{1,2}$ to the total population in the gain and absorber section, and $d_{1,2}$ to the corresponding spin inversion. Restricting to the fixed point subspace of the symmetry, now given by $F_{y}=0$, $d_{1}=0$ and $d_{2}=0$, and setting $F=F_x,$ $G=D_1$ and $Q=D_2$, we obtain
\begin{align}
2 F^\prime & =\left((1+i\alpha)G-(1+i\beta)Q-1\right)F \nonumber\\
& - \left(\varepsilon_{a}+i\varepsilon_{p}\right)F +\kappa e^{i\varphi}F(\cdot-\tau), \label{eq:Fred}\\
G^\prime & =\gamma_{G}\left(A-G\left(1+\left|F\right|^{2}\right)\right), \label{eq:Gred}\\
Q^\prime & =\gamma_{Q}\left(B-Q\left(1+a\left|F\right|^{2}\right)\right). \label{eq:Qred}
\end{align}
Notably, considering alternatively $F_{x}=0$, $d_{1}=0,$ and $d_{2}=0$, and setting $F=F_y,$ $G=D_1$ and $Q=D_2$ leads to the same equation with the opposite sign of $\varepsilon_p$.

In the absence of delayed feedback, system \eqref{eq:Fred}--\eqref{eq:Qred} reduces to the well-known Yamada model of a laser with SA \cite{Yamada1993}
\begin{align}
I^{\prime} & = \left[G - Q - 1 - \varepsilon_{a}\right]I, \label{eq:F-def-1-1-1} \\
G^{\prime} & = \gamma_{G}(A - G(1 + I)), \label{eq:G-def-1-1-1} \\
Q^{\prime} & = \gamma_{Q}(B - Q(1 + aI)), \label{eq:Q-def-1-1-1}
\end{align}
with the additional amplitude anisotropy term $\varepsilon_a$. The excitability properties of this model have been studied in much detail \cite{Dubbeldam1999}, and the mechanism of how purely intensity mediated delayed self-feedback leads to regenerative self-pulsation has been studied theoretically  \cite{Krauskopf2011,Terrien2018,Ruschel2020limits} and observed experimentally \cite{Terrien2017a,Terrien2018a}.

\begin{figure}[!]
\includegraphics[width=\columnwidth]{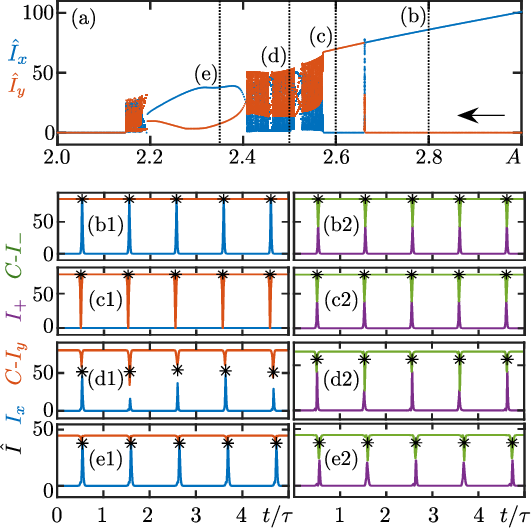}
\caption{Regenerative pulse trains of system \eqref{eq:F-def}--\eqref{eq:Q-def} for $\varepsilon_a=0.0$ and $\varepsilon_p=0.3$ as a function of the pump strength parameter $A$. (a) Parameter down-sweap in $A$ showing peak total intensity of the linear polarization components $\hat I_x$ (blue dots) and $\hat I_y$ (orange dots). (b)--(e) Representative time series at the $A$-values indicated in panel~(a), showing the component-wise intensity of pulse trains with fixed linear polarization angle in the linear basis in panels~(b1)--(e1) and in the circular basis in panel~(b2)--(e2). For illustration purposes, the $I_y$-component and the $I_-$-component are shown as the complements $C-I_y$ and $C-I_-$, where $C=\max_t (I_+(t)) + \max_t (I_-(t))$ is the sum of component-wise maximum peaks. Other parameters are as in Table \ref{tab:1}. \label{fig:bifurcationAepsp01}}
\end{figure}

\begin{table}[!]
\centering
\scalebox{1.0}{
\begin{tabular}{|c|c|c|c|c|c|c|}
\hline
$\alpha$ & $\beta $ & $\gamma_G$ & $\gamma_Q$ & $\delta_G$ & $\delta_Q$ & \\
\hline
$2.0$ & $0.5$ & $0.01$ & $0.01$ & $0.1$ & $0.1$ & \\
\hline\hline
$A$ & $B$ & $a$ & $\kappa$ & $\tau$ &$\varepsilon_a$ &$\varepsilon_p$ \\
\hline
$2.5$ & $2.0$ & $10.0$ & $0.2$ & $200$ & $[-0.1, 0.15]$ &$[0, 0.7]$\\
\hline
\end{tabular}
}
\caption{Model parameters and their values.}
\label{tab:1}
\end{table}

This analysis suggests the existence of purely $x$-polarized and purely $y$-polarized regenerative pulse trains of system \eqref{eq:F-def}--\eqref{eq:Q-def}. Figure~\ref{fig:bifurcationAepsp01} shows that they indeed exist, but that this depends on the value of the pump parameter $A$ of the gain section. Panel~(a) shows a down-sweep in $A$ of the peak total intensity of purely $x$-polarized periodic pulse trains in the feedback induced oscillatory regime $2.0 \leq A\leq 3.0$ and fixed phase anisotropy $\varepsilon_p = 0.3$ with $\varepsilon_a = 0$; see Table~\ref{tab:1} for the values of the other parameters. Figure~\ref{fig:bifurcationAepsp01}(b)--(e) shows representative time series at the selected values of $A$ that are indicated in panel~(a). Here, panels~(b1)--(e1) show the dynamics in the linear polarization basis, while panels~(b2)--(e2) present the same data in the circular polarization basis. In panel~(b1), the pulse train is purely $x$-polarized, with negligible intensity in the $y$ component; in panel~(c1), the polarization is fully rotated to the $y$ direction; in panel~(d1), we observe a weak but nonzero modulation of the peak pulse intensity, suggesting quasiperiodic behavior, and panel~(e1) shows amplitude modulation of both components, indicating the emergence of a mixed polarization state. The circular polarization components in panels~(b2)--(e2) provide additional insight into the underlying symmetry of the dynamics. In particular, panel~(e2) exhibits regenerative pulses consistent with a spatio-temporal symmetry of the form $F_+(t) = F_-(t - T/2)$, where $T$ is the temporal period of the pulse train. This behavior is characteristic of $\mathbb{Z}_2$-equivariant systems undergoing period-doubling bifurcations. In what follows, we investigate the effect of amplitude and phase anisotropy on the existence, stability and bifurcations of the different types of polarized pulse trains.

\section{Bifurcation analysis of strongly resonant locked periodic pulse trains in the $(\varepsilon_a,\varepsilon_p)$-plane}
\label{sec:2ParBifAna}

It was found in \cite{ruschel2025regenerative} that periodic pulse trains destabilize at torus bifurcations --- or, equivalently, Neimark-Sacker bifurcations of the corresponding Poincar{\'e} map --- where one finds the emergence of an invariant torus. The dynamics on the torus is either quasiperiodic or (frequency) locked. In the latter case, one observes locked periodic orbits (here, locked periodic pulse trains) in resonance regions that emerge from codimension-two resonance points on a torus bifurcation curve, where the rotation number\footnote{Recall that a rational rotation number $p\!:\!q$ of a locked periodic orbit encodes that the periodic orbit exhibits $q$ revolutions along `the parallel' and $p$ revolutions along `the meridian' of the torus before closing up on itself. When the rotation number is irrational, on the other hand, there is no closed trajetory on the torus.} is rational, that is, of the form $p\!:\!q$ for some $p,q\in\mathbb{N}$. The boundary of a $p\!:\!q$ resonance region, which has a cone-like shape and is also known as an Arnold tongue, is generally formed near the torus bifurcation by two curves of fold bifurcations of the locked periodic orbits (that is, periodic pulse trains). This statement holds when $q \geq 5$, and resonance tongues become small and very narrow quickly with increasing $q$. Hence, lower resonances are the important ones from the application perspective, especially the so-called strong resonances with $q = 1,2,3,4$, which involve other bifurcation curves; in particular, the region of $1\!:\!2$ locking is bounded by curves of period-doubling bifurcations; see, for example, \cite{kuznetsov2023two} for more details.

Torus bifurcations and associated resonance tongues are inherently a two-parameter phenomenon, and this motivates the investigation of the bifurcations of periodic pulse trains with fixed polarization angle with respect to two parameters. Since we have identified the anisotropy parameters $\varepsilon_a$ and $\varepsilon_p$ as playing a role in the emergence of locked periodic pulse trains, we focus our attention on the corresponding bifurcation structure in the ($\varepsilon_a,\varepsilon_p$)-parameter plane. More specifically, we consider the torus bifurcation curves $\textbf{T}_x$ and $\textbf{T}_y$ of the purely $x$-polarised and purely $y$- polarised pulse trains, their points of strong resonances $1\!:\!2$, $1\!:\!3$ and $1\!:\!4$ and the corresponding regions of locked pulse trains. These objects are found by means of a numerical bifurcation analysis performed with the continuation software package DDE-Biftool \cite{Engelborghs2002,Sieber2014} for \textsc{Matlab} or GNU Octave. 

\begin{figure}[!]
        \vspace*{2mm}
        \includegraphics[width=\columnwidth]{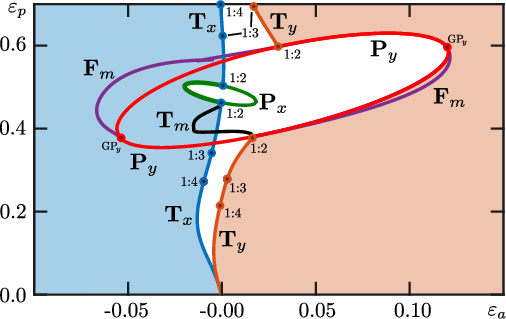}
\centering{}\caption{Two-parameter bifurcation diagram of system \eqref{eq:F-def}--\eqref{eq:Q-def} in the $(\varepsilon_{a},\varepsilon_{p})$-plane of pulse trains with fixed polarization angle, showing curves of torus bifurcation $\textbf{T}_x$ (blue), $\textbf{T}_m$ (black), and $\textbf{T}_y$ (orange), of fold bifurcations $\textbf{F}_m$ (purple), and of period-doubling bifurcations $\textbf{P}_x$ (green) and $\textbf{P}_y$ (red) of the purely x-polarised pulse train, mixed, and purely y-polarised pulse trains, respectively, as indicated by subscripts; points of strong resonances on $\textbf{T}_x$ (blue)  and $\textbf{T}_y$ are marked and labelled; points of degenerate period-doubling bifurcation GP$_y$ are marked and labelled. Purely x-polarised and purely y-polarised periodic pulse trains are locally stable in the blue and orange shaded regions; in the white region one may find pulse trains with mixed polarisation angle. Other parameters are as in Table \ref{tab:1}. \label{fig:2ParBifDiag}}
\end{figure}

Figure~\ref{fig:2ParBifDiag} shows the resulting bifurcation structure and stability information of the different types of periodic pulse trains in the $(\varepsilon_{a},\varepsilon_{p})$-plane over the ranges $-0.10\leq \varepsilon_a\leq 0.15$ and $0.00\leq \varepsilon_p \leq 0.65$. Purely $x$-polarized pulse trains are stable in the large region on the left (blue shading); they destabilize for increasing $\varepsilon_{a}$ at the torus bifurcation curve $\textbf{T}_x$ and at the left arc of the period-doubling bifurcation curve $\textbf{P}_x$, which connects two $1\!:\!2$ resonance points on $\textbf{T}_x$. Purely $y$-polarized pulse trains are stable in the large region on the right (orange shading); similarly, they destabilize for decreasing $\varepsilon_{a}$ at the torus bifurcation curve $\textbf{T}_y$ and the right arc of the period-doubling curve $\textbf{P}_y$ connecting two $1\!:\!2$ resonance points. Other points of strong resonances are shown on the curves $\textbf{T}_x$ and $\textbf{T}_y$, and we find that, for increasing $\varepsilon_{p}$, the rotation number increases up to the respective first $1\!:\!2$ resonance point, and then decreases from the respective second $1\!:\!2$ resonance point. In addition, we find in Fig.~\ref{fig:2ParBifDiag} a curve $\textbf{T}_m$ of torus bifurcation of the period-doubled pulse trains, which connects a $1\!:\!2$ resonance point on $\textbf{T}_x$ and a $1\!:\!2$ resonance point on $\textbf{T}_y$. Moreover, there are two curves $\textbf{F}_m$ of fold bifurcations of period-doubled pulse trains that each connect a degenerate period-doubling bifurcation points GF$_y$ on the curve $\textbf{P}_y$ with a $1\!:\!2$ resonance point on $\textbf{T}_y$.

The curves $\textbf{T}_m$ and $\textbf{T}_x$ form (part of the) boundary of where in the intermediate region (white) one finds stable periodic pulse trains with constant linear polarization angle that are neither purely $x$-polarized nor purely $y$-polarised. We refer to this tpye of periodic pulse train as having \emph{mixed linear polarization angle}, or simply \emph{mixed} for short. To explain the emergence of the mixed polarization angle regime and illustrate the possible dynamics and transitions between of different types of periodic pulse trains, we now consider one-parameter bifurcation diagrams for chosen fixed values of $\varepsilon_p$-value. Here, we focus attention on the dynamics in the immediate proximity of the  $1\!:\!2$, $1\!:\!3$ and $1\!:\!4$ resonance points along the lower part of the curve $\textbf{T}_y$ in Fig.~\ref{fig:2ParBifDiag}. 

\subsection*{$1\!:\!2$ locked pulse trains}

\begin{figure}[!]
        \vspace*{2mm}
	\includegraphics[width=\columnwidth]{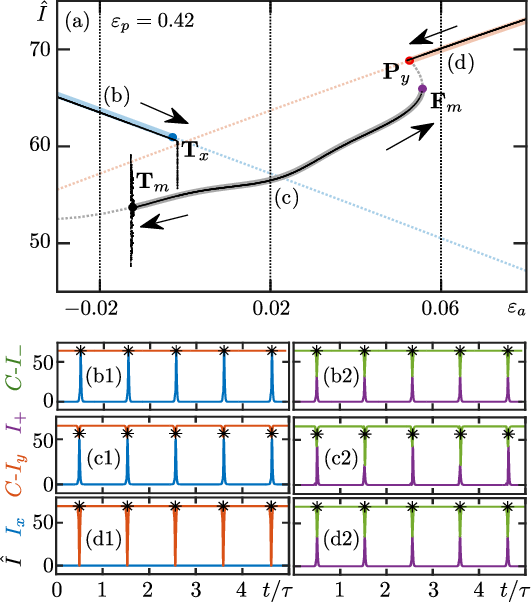}
	\centering{}\caption{Characterization of periodic pulse trains with fixed polarization angle of system \eqref{eq:F-def}--\eqref{eq:Q-def} during the transition through a $1\!:\!2$ strong resonance for fixed $\varepsilon_p=0.42$. (a) Shown in terms of peak total intensity $\hat{I}$ are: branches obtained by continuation of purely $x$-polarized (blue), mixed (gray), and purely $y$-polarized (orange) pulse trains (stable when solid; unstable when dashed), with the bifurcation points $\textbf{T}_m$ (black), $\textbf{T}_x$ (blue), $\textbf{P}_y$ (orange) and $\textbf{F}_m$ (purple); together with points (black) obtained from an up-sweep and a down-sweap in $\varepsilon_a$, as indicated by arrows. (b)--(d) Representative time series at the $\varepsilon_a$-values $-0.02$, $0.02$ and $0.06$ indicated in panel~(a), showing the component-wise intensity of pulse trains with fixed linear polarization angle, in the linear basis in panels~(b1)--(d1) and the circular basis in panel~(b2)--(d2); as in Fig.~\ref{fig:bifurcationAepsp01}(b)--(e), the representation is in terms of $I_x$, $I_+$ and the complements $C-I_y$ and $C-I_-$. Other parameters are as in Table \ref{tab:1}. 
\label{fig:2ParBifDiagSlice1}}
\end{figure}

We now fix $\varepsilon_p=0.42$ and consider the one-parameter slice with $-0.03\leq \varepsilon_a\leq0.08$, which cuts through the lower curves $\textbf{T}_m$, $\textbf{T}_x$, $\textbf{P}_y$ and $\textbf{F}_m$ in Fig.~\ref{fig:2ParBifDiag}. The resulting one-parameter bifurcation diagram is shown in Fig.~\ref{fig:2ParBifDiagSlice1}(a) in terms of peak total intensity $\hat{I}$. It shows branches of periodic orbits obtained by continuation with their points of bifurcations, together with points obtained by parameter sweeping, where the attractor is `tracked' by using the last time segment of length $\tau$ as the initial condition for the numerical integration of Eqs.~\eqref{eq:F-def}--\eqref{eq:Q-def} at the next (small) parameter increment. Specifically, panel~(a) shows simultaneously the up-sweep starting from the purely $x$-polarized periodic pulse train at $\varepsilon_a=-0.03$, and the down-sweep from the purely $y$-polarized periodic pulse train at $\varepsilon_a=0.08$, as indicated by arrows. Panels~(b)--(d) show  representative time series at the $\varepsilon_a$-values indicated in panel~(a), in terms of the linear basis in panels~(b1)--(d1) and the circular basis in panel~(b2)--(d2).

We observe two distinct transitions in Fig.~\ref{fig:2ParBifDiagSlice1}: between a purely $x$-polarized and a periodic pulse train with a mixed polarization angle, and between this mixed and a purely $y$-polarized periodic pulse train, each featuring a hysteresis loop. In the first transtion, there is a sudden switch for increasing $\varepsilon_a$ from a purely $x$-polarized to a mixed pulse train at the torus bifurcation $\textbf{T}_x$, due to the fact that this torus bifurcation is subcritical. For decreasing $\varepsilon_a$, we find a very similar scenario with a quite sudden transition from a mixed to a purely $x$-polarized periodic pulse train at the subcritical torus bifurcation $\textbf{T}_m$; we remark that the small $\varepsilon_a$-range with seemingly rapidly growing quasiperiodic behavior is due to very long transients in light of the only weak instability of the $x$-polarized periodic orbit past $\textbf{T}_m$. Representatives of the two different types of periodic pulse trains are shown in panel~(b) of Fig.~\ref{fig:2ParBifDiagSlice1} for $\varepsilon_a=-0.02$ and in panel~(c) for $\varepsilon_a =0.02$. Although the two time-series appear to be almost identical in the linear polarization basis at first glance, they are qualitatively different. While the peak total intensity of the mixed pulses in panel~(c1) still has a dominant contribution from the $x$-component $I_x$, there is a small contribution of the $y$-component $I_y$ to the peak total intensity. Panel~(c2) shows the same time series in the circular polarization basis, where one clearly observes equidistant antiphase pulsing dynamics with exactly two maximum peak values of $I_+$ and $I_-$. The fact that the peak total intensity remains the same hints at the fact that this pulse train exhibits the spatio-temporal symmetry $F_+(t)=F_{-}(t-T/2)$ (where $T \approx 2\tau$ is the underlying period); this is easily verified by (mentally) shifting the two time series with respect to one another by $T/2\approx\tau$ units of time. This type of behavior is known to arise from a period-doubling bifurcation in a $\mathbb{Z}_2$-equivariant dynamical system \cite{golubitsky2003symmetry}, such as system \eqref{eq:F-def}--\eqref{eq:Q-def}.

When following the branch of mixed periodic pulse trains in Fig.~\ref{fig:2ParBifDiagSlice1}(a) towards larger values of $\varepsilon_a$, we observe that it disappears in the second transitions with a hysteresis loop. Specifically, for increasing $\varepsilon_a$, there is a sudden transtion at the fold bifurcation $\textbf{F}_m$ to the the purely $y$-polarized branch. This bifurcation is closely related to the nearby period-doubling bifurcation $\textbf{P}_x$, where there is a sudden transition form a purely $y$-polarized to a mixed periodic pulse train for decreasing $\varepsilon_a$; hence, $\textbf{P}_x$ is subcritical, as is expected from the existence of the nearby fold bifurcation $\textbf{F}_m$. The representative time series of the purely $y$-polarized pulse train in Fig.~\ref{fig:2ParBifDiagSlice1}(d1) and~(d2) shows that the overall pulse intensity is now indeed entirely due to $I_y$, while the circluar components $I_+$ and $I_-$ are again equal, as is the case for the purely $x$-polarized pulse train in panel~(b2).

\subsection*{$1\!:\!3$ locked pulse trains}

\begin{figure}[!]
	\includegraphics[width=\columnwidth]{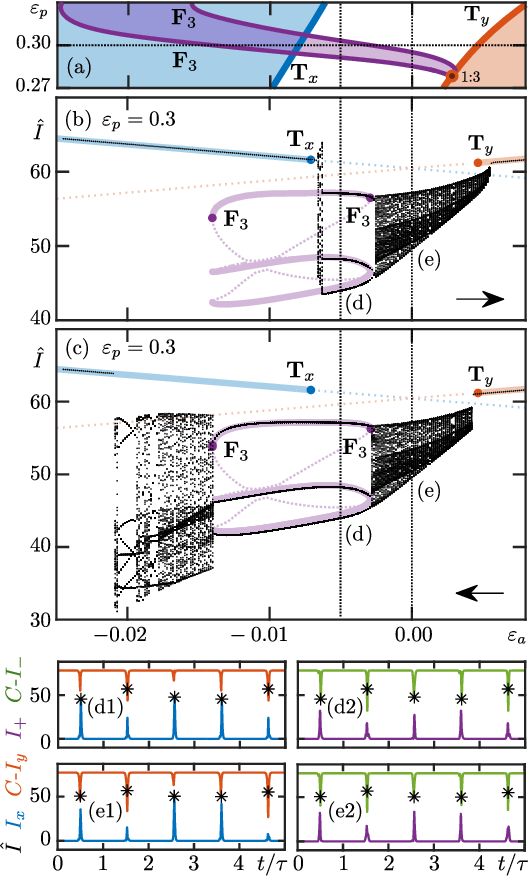}
	\centering{}\caption{Characterization of periodic pulse trains with fixed polarization angle of system \eqref{eq:F-def}--\eqref{eq:Q-def} during the transition through a $1\!:\!3$ strong resonance for fixed $\varepsilon_p=0.3$ with $-0.03 \leq \varepsilon_a \leq 0.08$. (a) Enlargement of the $(\varepsilon_a,\varepsilon_p)$-plane in Fig.~\ref{fig:2ParBifDiag} near the $1\!:\!3$ resonance point on the curve $\textbf{T}_y$; also shown is the pair of fold bifurcation curves $\textbf{F}_3$ (purple) that bound the region (purple shading) with $1\!:\!3$ locked periodic pulse trains. (b)--(c) Shown in terms of peak total intensity $\hat{I}$ are: branches obtained by continuation of purely $x$-polarized (blue), purely $y$-polarized (orange) and $1\!:\!3$ locked (purple) periodic pulse trains (stable when solid; unstable when dashed), with the bifurcation points $\textbf{T}_x$ (blue), $\textbf{T}_y$ (orange), and $\textbf{F}_3$ (purple); together with points (black) obtained from an up-sweep and a down-sweap in $\varepsilon_a$, as indicated by the arrows. (d)--(e) Representative time series at the $\varepsilon_a$-values $-0.005$ and $0.0$ indicated in panel~(a), showing the component-wise intensity of pulse trains with fixed linear polarization angle, in the linear basis in panels~(d1) and~(e1) and the circular basis in panel~(d2) and~(e2), represented by $I_x$, $I_+$ and the complements $C-I_y$ and $C-I_-$; see Fig.~\ref{fig:bifurcationAepsp01}. Other parameters are as in Table \ref{tab:1}.
\label{fig:2ParBifDiagSlice2}}
\end{figure}

We now illustrate in Fig.~\ref{fig:2ParBifDiagSlice2} what the transition between purely $x$-polarized and purely $y$-polarized periodic pulse trains looks like when the $1\!:\!3$ resonance region in the $(\varepsilon_a,\varepsilon_p)$-plane is crossed. Panel~(a) shows an enlargement of the parameter region near the $1\!:\!3$ resonance point on the torus bifurcation curve $\textbf{T}_y$, with the addition of the region with locally stable $1\!:\!3$ locked periodic orbits, which is bounded by a pair of fold bifurcation curves $\textbf{F}_3$. Note that the region of stable $1\!:\!3$ locked periodic orbits overlaps with that of stable purely $y$-polarized periodic orbits to the left if the torus bifurcation curve $\textbf{T}_x$, indicating multistability between these two types of periodic pulse trains.

To illustrate the dynamics inside and outside of the region of $1\!:\!3$ locking, we consider the one-parameter bifurcation diagram along the parameter slice for fixed $\varepsilon_p=0.3$ and $-0.03 \leq \varepsilon_a \leq 0.08$, which corresponds to the black dashed line in panel~(a). The continued branches of purely $x$-polarized, $1\!:\!3$ locked and purely $y$-polarized periodic orbits and their bifurcation points $\textbf{T}_x$, $\textbf{T}_y$ and $\textbf{F}_3$, are presented with a parameter up-sweep in panel~(b), and a parameter down-sweep in panel~(c) of Fig.~\ref{fig:2ParBifDiagSlice2}. Comparison of these two sweeping directions immediately confirms a large range of bistability associated with the switching between purely $x$-polarized, and $1\!:\!3$ locked periodic pulse trains.

Focusing first on the up-sweep in Fig.~\ref{fig:2ParBifDiagSlice2}(b), we observe that the purely $x$-polarized periodic pulse train is stable up to the point $\textbf{T}_x$. After a small $\varepsilon_a$-range with rapidly growing quasiperiodic behavior due to very long transients, the periodic pulse train settles on a stable $1\!:\!3$ locked periodic orbit, displaying three distinct values of pulse height. The corresponding $1\!:\!3$ locked periodic pulse train is shown in panels~(d1) and~(d2): it features equidistant pulses of three distinct levels of the linear polarizations $I_x$ and $I_y$ and circular polarizations $I_+$ and $I_-$, respectively, which keep repeating in the same pattern with an overall period of about $3\tau$. When further increasing $\varepsilon_a$, the $1\!:\!3$ locking region is exited when the fold bifurcation $\textbf{F}_3$ on the right is crossed; here, the stable $1\!:\!3$ locked periodic orbits meets its saddle counterpart and both disappear. The system then switches to dynamics on the stable torus that is now quasiperiodic or of very high period; the corresponding pulse train is shown in panels~(a1) and~(e2): the pulses are still equidistant but now have an amplitude modulation that does not repeat for all practical purposes. For increasing $\varepsilon_a$, the torus shrinks down in size and the system switches to a purely $y$-polarized periodic pulse train at $\varepsilon_a\approx0.015$; this happens at a `fold bifurcation of tori' near the actual torus bifurcation $\textbf{T}_y$, which is actually subcritical.

The subcriticality of $\textbf{T}_y$ is confirmed by the down-sweep starting from a purely $y$-polarized periodic orbit: for decreasing $\varepsilon_a$, the system suddenly switches to the stable torus only at $\textbf{T}_y$, thus, creating a small hysteresis loop. The torus is stable down to $\varepsilon_a \approx -0.022$; here, it disappear in a boundary crisis bifurcation, leading to a sudden transition to the stable purely $x$-polarized periodic orbit for lower values of $\varepsilon_a$. Note that, in the process of tracking the dynamics on the stable torus for decreasing $\varepsilon_a$, the $1\!:\!3$ locking region is entered at the right fold bifurcation $\textbf{F}_3$, traversed in its entirety, and then exited at the left fold bifurcation $\textbf{F}_3$. To the left of the $1\!:\!3$ locking region one can observe a couple of larger $\varepsilon_a$-ranges where the dynamics is $p\!:\!q$ locked with fairly low $q$.

\subsection*{$1\!:\!4$ locked pulse trains}

\begin{figure}[!]
        \vspace*{2mm}
	\includegraphics[width=\columnwidth]{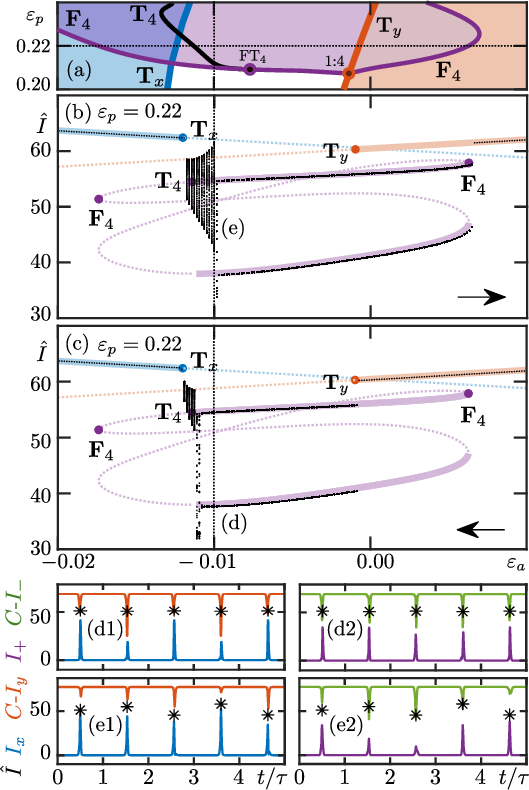}
        \caption{Characterization of periodic pulse trains with fixed polarization angle of system \eqref{eq:F-def}--\eqref{eq:Q-def} during the transition through a $1\!:\!4$ strong resonance for fixed $\varepsilon_p=0.22$ with $-0.02\leq \varepsilon_a\leq0.05$, presented in the style of Fig.~\ref{fig:2ParBifDiagSlice3}. (a) Enlargement of the $(\varepsilon_a,\varepsilon_p)$-plane near the $1\!:\!4$ resonance point on $\textbf{T}_y$, with the pair of fold bifurcation curves $\textbf{F}_4$ (purple) bounding the region (purple shading) with $1\!:\!4$ locked periodic pulse trains. The $1\!:\!4$ resonant point is marked and labelled. Also shown is a torus bifurcation curve \textbf{T}$_4$ emanating from the leftmost curve \textbf{F}$_4$ at point labelled FT$_4$. (b)--(c) Branches of purely $x$-polarized (blue), purely $y$-polarized (orange) and $1\!:\!4$ locked (purple) periodic pulse trains (stable when solid; unstable when dashed), with the bifurcation points $\textbf{T}_x$ (blue), $\textbf{T}_y$ (orange) and $\textbf{F}_4$ (purple), as well as the torus bifurcation point $\textbf{T}_4$ (black) of the $1\!:\!4$ locked periodic orbit shown together with points (black) obtained from an up-sweep and a down-sweep in $\varepsilon_a$, as indicated by the arrows. (d)--(e) Representative time series at $\varepsilon_a = -0.01$ of the $1\!:\!4$ locked periodic pulse train and the simultaneously existing quasiperiodic pulse train, respectively, shown in terms of $I_x$, $I_+$, $C-I_y$ and $C-I_-$; see Fig.~\ref{fig:bifurcationAepsp01}. Other parameters are as in Table \ref{tab:1}.
\label{fig:2ParBifDiagSlice3}}
\end{figure}

Figure~\ref{fig:2ParBifDiagSlice3} illustrates in the same spirit that one may also encounter strong $1\!:\!4$ locking in the transition between purely $x$-polarized and purely $y$-polarized pulse trains. The enlargement of the $(\varepsilon_a,\varepsilon_p)$-parameter near the $1\!:\!4$ resonant point on $\textbf{T}_y$ in panel (a) also shows a pair of fold bifurcation curves $\textbf{F}_4$, and a torus bifurcation curve $\textbf{T}_4$ that bound the region with stable $1\!:\!4$ locked periodic orbits. Both curves $\textbf{F}_4$ emanate from the $1\!:\!4$ resonant point on $\textbf{T}_y$, and the curve $\textbf{T}_4$ intersects the lower arc of the leftmost $\textbf{F}_4$ curve at a point FT$_4$ which corresponding to a $1\!:\!1$ resonant point on $\textbf{T}_4$ (labelled in a different convention to avoid confusion with resonant points along $\textbf{T}_y$). The one-parameter bifurcation diagram along the parameter slice for fixed $\varepsilon_p=0.22$ is shown in panel~(b) with an up-sweep, and in panel~(c) with a down-sweep of $-0.03 \leq \varepsilon_a \leq 0.08$; here, we show the branches of purely $x$-polarized and purely $y$-polarized periodic pulse trains with the points $\textbf{T}_x$, $\textbf{T}_y$, as well as the branch of $1\!:\!4$ locked periodic pulse trains with its bifurcation points $\textbf{F}_4$ and $\textbf{T}_4$.

When starting from the stable purely $x$-polarized pulse train and increasing $\varepsilon_a$, as in Fig.~\ref{fig:2ParBifDiagSlice3}(b), one observes a sudden transition at the point $\textbf{T}_x$ to a stable torus that emerges from this torus bifurcation, which is again subcritial with a `fold of tori' just to its left; the dynamics on the torus is quasiperidic or of very high period. As $\varepsilon_a$ is increased further, the stable torus disappears in a boundary crisis at $\varepsilon_a \approx -0.01$, where the system switches to a stable $1\!:\!4$ locked periodic orbit. In fact, there is a small region of bistability between these two types of pulse trains; in particular, they coexist for $\varepsilon_a=-0.01$, and panels~(d) and~(e) show the corresponding time series of the $1\!:\!4$ and quasiperiodic locked pulse trains, respectively. Notably, the family of $1\!:\!4$ locked periodic orbits is invariant under the spatio-temporal symmetry of the applying the exchange involution $\eta$ (which exchanges $I_x$ and $I_y$) and a time shift over $T/2$, where $T$ is again the total period (of about $4 \tau$ in this case). This is why the $1\!:\!4$ locked periodic orbit is represented by only two branches with values of the total peak intensity $\hat{I}$ in Fig.~\ref{fig:2ParBifDiagSlice3}(d1). However, the peak intensities of the circular polarization components $I_+$ and $I_-$ in panel~(d2) have four distinct peak intensities and, hence, clearly illustrate the $1\!:\!4$ locked nature of this pulse train. In contrast, on the simultaneously existing stable torus, the system exhibits pulse trains with non-repeating modulation of the peak total and component-wise intensities, as is shown in panels~(e1) and~(e2); this quasiperiodic dynamics gives rise to the many dots at $\varepsilon_a=-0.01$ in Fig.~\ref{fig:2ParBifDiagSlice3}(b). When further increasing $\varepsilon_a$ past the boundary crisis of the stable torus, the system follows the stable $1\!:\!4$ locked periodic orbit, until the right fold point $\textbf{F}_4$ is reached, where there is a sudden transition to stable $y$-polarized periodic pulse trains. 

The down-sweep in Fig.~\ref{fig:2ParBifDiagSlice3}(c) shows that the system switches back from stable purely $y$-polarized to $1\!:\!4$ locked periodic pulse trains only at the torus bifurcation point $\textbf{T}_y$, yielding a hysteresis loop over a quite large $\varepsilon_a$-range. The $1\!:\!4$ locked periodic orbit loses its stability at its torus bifurcation point $\textbf{T}_4$; this bifurcation is subcritical and there is a switch to the stable periodic torus, which is followed for decreasing $\varepsilon_a$ down to where it disappears in a `fold of tori' near the subcritical torus bifurcation point $\textbf{T}_x$. Subseqently, the system switches to following the branch of stable purely $x$-polarized periodic pulse trains. The associated hysteresis loop suggest a complicated bifurcation structure of invariant tori near the points $\textbf{T}_x$ and $\textbf{T}_4$, the details of which are beyond the scope of this paper.

\section{Conclusions}
\label{sec:concl}

We investigated the emergence and structure of polarization-locked regenerative pulse trains in a spin-flip model for an excitable microlaser with a saturable absorber and delayed optical feedback. More specifically, we performed a numerical bifurcation analysis to determine how amplitude and phase anisotropies, represented by the parameters $\varepsilon_a$ and $\varepsilon_p$, affect the polarization dynamics of self-sustained pulses. Our focus was on strong locking of the peak intensity dynamics to 2, 3, and 4 times the self-pulsation frequency. We identified the regions in the $(\varepsilon_a,\varepsilon_p)$-plane where these special types of pulse trains can be found; namely, we identified and continued their boundary curves of period-doubling or fold bifurcations that emerge from corresponding codimension-two strong resonance points on a curves of torus bifurcation. One-parameter bifurcation diagrams in $\varepsilon_a$, for suitable values of $\varepsilon_p$, showed that $1\!:\!2$, $1\!:\!3$ and $1\!:\!2$ locked pulse trains are encountered over considerable $\varepsilon_a$-ranges --- as part of the transition between purely $x$-polarized and purely $y$-polarized periodic pulse trains.

These findings highlight the dynamical richness introduced by even weak polarization coupling in microlasers with delayed self-coupling. From a nonlinear dynamics point of view, the interaction between internal polarization modes and external delay creates a structured yet complex landscape of resonances, quasi-periodicity, and mode competition. Such phenomena are not only of theoretical interest but also directly relevant for photonic devices designed to emulate neural behavior.

From an applied perspective, our analysis constitutes a step towards exploiting polarization-encoded spiking patterns in photonic neural circuits, as envisioned in recent works. On the one hand, the existence of robust strongly polarization-locked pulse trains suggests a way to encode information in the polarization state of light pulses for neuromorphic computing.  For instance, a linearly polarized pulse in a single polarization component alone could represent a spike, and the switch to the orthogonal polarization component could be used as a logical operation. On the other hand, our analysis suggests that, in the presence of even vanishingly small anisotropy of the microlaser, pulse propagation in only one polarization component, or in both linear polarization components independently, as theoretically proposed in \cite{Zhang2018,ZhangOL19}, might not be robust without a strategy to compensate for the anisotropy cross-coupling, for example, by additional polarization filtering of the feedback. These insights have practical implications for polarization-encoded spike processing in neuromorphic photonics. For example, locking to specific polarization sequences may be used to implement multilevel spike coding or polarization-selective logic gates in integrated optical hardware. 

Interesting directions for future work, in this context, include the exploration of the impact of noise on the switching between locked states; spatially extended effects of switching, for example, in coupled microlaser arrays; and the experimental realization of higher-order polarization locking under engineered feedback conditions.

\section*{Acknowledgments}
S.R. thanks Jan Sieber for advice regarding the numerical implementation in DDE-Biftool, and acknowledges support by UKRI Grant No. EP/Y027531/1.

\section*{Data Availability Statement}
The data that support the findings of this study are available from the authors upon reasonable request.





\end{document}